\documentclass[12pt]{article}

\usepackage{epsfig}

\newcommand{\thetabao}{\Delta\tilde{\theta}_{\rm B}}
\newcommand{\dlbao}{\Delta\tilde{\lambda}_{\rm B}}
\newcommand{\dzbao}{\Delta \tilde{z}_{\rm B}}
\newcommand{\Qsn}{Q_{\rm SN}}
\newcommand{\Qnic}{Q_{\rm Ni}}
\newcommand{\QH}{Q_{\rm H}}
\newcommand{\lla}{\lambda_{\rm Ly\alpha}}
\newcommand{\llam}{\tilde{\lambda}_{\rm Ly\alpha}}
\newcommand{\lsnm}{\tilde{\lambda}_{\rm SN}}
\newcommand{\lya}{Ly$\alpha\;$}
\newcommand{\Adet}{A_{\rm det}}

\title{Dimensionless constants and cosmological measurements}
\author{J.\,Rich \\ IRFU-SPP, CEA Saclay \\ 91191 Gif-sur-Yvette, 
France}

\begin{document} 

\maketitle                         

\begin{abstract}
The laws of physics have a set of 
fundamental constants, and it is generally admitted
that only dimensionless combinations of constants
have physical significance.
These combinations include 
the electromagnetic and gravitational 
fine structure constants, $\alpha=e^2/4\pi\epsilon_0\hbar c$ and 
$\alpha_G=Gm_p^2/\hbar c$, along with
the ratios of elementary-particles masses.
Cosmological measurements clearly depend on the values of these
constants in the past and can therefore give
information on their time dependence
if the effects of time-varying constants can be separated
from the effects of cosmological parameters.
The latter can be eliminated by using pairs of redundant measurements
and here we show how such pairs
conspire to give
information only on dimensionless combinations of constants.
Among other possibilities, we will use  distance measurements  based on
Baryon Acoustic Oscillations (BAO) 
and on  type Ia supernova.
The fact that 
measurements yield information only on
dimensionless combinations 
is traced to the fact that 
distances between co-moving points
expand following
the same   function of time that governs the
redshift of photon wavelengths.
\end{abstract}

\section{Introduction} 

\setlength{\baselineskip}{4.0ex}

Cosmological observations 
of high-redshift objects tell us about the conditions in the
distant past and it is natural to try to use them to
see if the physical laws have changed over time.
One simple type of change would involve time variations of
the fundamental constants  (hereafter TVOFCs) \cite{uzan}.
The most famous searches for TVOFC 
are those for time variations of the
fine-structure constant, $\alpha$, that are performed by looking for
a redshift dependence in the splitting
of atomic or molecular spectral lines.  
Current results using atomic fine-structure splittings
give controversial evidence for \cite{murphy} or against \cite{petitjean}
such a variation at the level of $\Delta\alpha/\alpha\sim10^{-5}$
over $\sim10^{10}$yr.
This sort of study makes use of redundant cosmological 
information:  a spectral doublet has two lines that can be used to
measure the redshift.  
Because the cosmological redshift is achromatic,
any difference in the two redshifts
could be explained by a different fundamental 
constants at the time when the photons were produced,
leading to a splitting different from its present value.

Another suggestion for TVOFCs comes from the observed
fluxes of high-redshift type-Ia supernovae (SNIa) 
\cite{Riess98,Perlmutter99,sndistances,union}.
These events have fluxes about 40\% smaller than what
would be expected in a universe where the expansion was
being decelerated by the ordinary gravitational attraction 
of a critical density of matter. 
The generally accepted explanation for
this low flux is that the universe contains
dark-energy that accelerates the expansion, causing
the  supernovae to be
about 20\% further from us (at a given redshift) than 
expected.  
An alternative explanation of the faintness of distant SNIa
would be that in the past they were less luminous than 
those that explode now.  This might be simply due to the astrophysical
conditions changing over time.  
A more interesting possibility would be that
the fundamental constants governing SNIa luminosities were
different in the past.
We will see that SNIa luminosities depend strongly on the 
gravitational constant, $G$, at the moment of the explosion.
Of course $G$ also determines the expansion of the universe
so the supernova flux we can expect to receive depends also on
the entire history of this constant, $G(t)$, between
explosion and detection.
Analyses of SNIa data taking into account both supernova
luminosities and light propagation have been performed
\cite{riazuelo,garciaberro,prosper,wei} yielding limits on the time variation of $G$,
generally supposing  that all other fundamental constants
are time-independent.

The fact that one can envision both TVOFC and dark-energy interpretations
of SNIa measurements illustrates the fact that
a given type of observation can yield cosmological energy densities
or fundamental constants but not both.
This problem can be eliminated by comparing with 
a second distance measurement \cite{bassett}. As an example, we use here
measurements  provided by the 
Baryonic Acoustic Oscillation
(BAO) ``standard ruler'' consisting of  a peak in the
matter two-point correlation function at a co-moving distance
of $\sim150~{\rm Mpc}$.
Distances to objects at a given redshift determined with BAO
can be directly compared to distances deduced from the fluxes
of SNIa  ``standard candles''.
This will allow us to

\begin{itemize}
\item distinguish unambiguously between explanations
of the low SN flux that are based on dark-energy from those
based on TVOFC, 
or, more generally, on any physics causing ancient supernovae
to be less luminous than modern ones.
\end{itemize}
and
\begin{itemize}
\item illustrate why only dimensionless combinations of 
fundamental constants can be studied with cosmological measurements.
\end{itemize}
Since it is well known that 
SNIa distances \cite{sndistances,union}
agree with BAO distances \cite{Chuangref,Seoref,Xuref,Anderson2013},
it will come as no surprise that 
we will not find  evidence for TVOFC.
Furthermore, the precision of current cosmological
measurements (percent level) combined with 
astrophysical uncertainties will make our limits on TVOFC
uncompetitive with those found by other techniques.
It is therefore the  
second
point
which is the most important, i.e.  placing cosmological measurements
in the framework of ``dimensionless cosmology'' \cite{narimani}.
The way that pairs of measurements  conspire to yield
only information on dimensionless combinations will turn out
to be simple but not entirely trivial:  the BAO ruler expands
with time in the same way as photon wavelengths.  In other
words, both the BAO ruler and photon wavelengths were smaller
in the past by the same factor $(1+z)$ where $z$ is the redshift.
To the extent that physics respects this rule, we will be 
condemned to study only dimensionless  combinations of
constants, a fact that
may come
as an embarrassment for advocates of non-standard theories
where, for example, the speed of light is allowed to vary
with time \cite{Moffat}.

For measurements performed over a short period of time, 
the fact that only dimensionless combinations influence measurements
follows from
simple dimensionless analysis \cite{Cook,Turneaure,rich}.  
For example, if we measure the speed of light by measuring
the time for a photon to travel the length of a rigid rod,
we actually measure the number, $N_t$, of periods of some
standard clock required for the trip.  If we use an atomic clock,
the period will be a well-defined function of 
fundamental constants.  The length of the rod is some number, $N_x$,
of inter-atomic spacings and this spacing can be assumed to depend
on the fundamental constants in the same ways as the Bohr radius,
$a_B=\hbar/\alpha m_ec$, being the only length
formed from the relevant constants $(\hbar,e,m_e)$ \cite{Turneaure}. 
Before the measurement is performed,
the only meaningful question we can ask is what 
will  $N_x/N_T$ turn out to be? 
Being dimensionless itself,  $N_x/N_T$  can
then only depend on a dimensionless combination of fundamental constants.
For example, the use of a hydrogen maser as a clock gives
$N_T/N_x\propto g\alpha^3(m_e/m_p)$ where $g$ is the
electron gyromagnetic ratio,  and $m_e$ and
$m_p$ are  the electron and proton masses.

Similarly it can be shown \cite{rich} that local measurements
of the stability of 
planetary orbits are sensitive not to time variations of $G$ but
rather to those of the ``gravitational fine-structure constant''
\begin{equation}
\alpha_G=\frac{Gm_p^2}{\hbar c} = 5.9\times10^{-39}
\end{equation}
In particular, monitoring orbital parameters with a radar
and atomic clock give limits on the time variation
of $\alpha_G g\alpha^4(m_e/m_p)$.

These two examples illustrate how the effect of fundamental
constants on local measurements comes both through the phenomenon
being observed and  through the measurement apparatus. 
The two effects combine to give sensitivity only to dimensionless combinations.
Showing that \emph{cosmological} measurements depend only on dimensionless
fundamental constants is more delicate because they are much more
complicated than the above examples.
Care must be taken to model the influence of fundamental constants
on all relevant aspects: the astrophysical phenomenon to be
observed, the propagation of photons from the source to observer,
and the act of observation.
In what follows, we will show how dimensionality enters into
two fundamental types of cosmological measurements, those of distances
and of expansion rates

\section{Distances}

\subsection{Type Ia Supernovae}

The most precise method for measuring distances to galaxies
is to measure the photon flux from a type Ia supernova in the galaxy.
We therefore start by discussing how the fundamental constants
enter SNIa observations.
The production of visible light in SNIa is
believed to be due to the energy released in the beta decays
of  the ${\rm ^{56}Ni}$ produced in the nuclear reactions that
drive the explosion.
The amount of  ${\rm ^{56}Ni}$ is of order the Chandrasekhar mass
which, apart from factors of order unity 
related to the relative numbers of neutrons
and protons in the pre-explosion star, is given by
\begin{equation}
M_{\rm ch} \sim \alpha_G^{-3/2}m_p \sim 2.2\times 10^{57}m_p
\end{equation}
The total energy available for visible photon production is then
the number of nickel nuclei times the energy release per decay:
\begin{equation}
\Qsn \sim \frac{1}{56\alpha_G^{3/2}} \times \Qnic
\end{equation}
where $\Qnic$ is the energy release in the beta decay
sequence ${\rm ^{56}Ni}$ ${\rm \rightarrow^{56}Co}$ 
${\rm \rightarrow^{56}Fe}$ ,
a quantity that depends of the neutron-proton mass difference and the
relative binding energies of   ${\rm ^{56}Ni}$, ${\rm ^{56}Co}$
and  ${\rm ^{56}Fe}$.
Presumably, $\Qnic$ is determined by the 
relevant fundamental constants,
$\Lambda_{\rm QCD}$, $\alpha$, and quark masses.
It  is not presently possible to give a useful formula for $\Qnic$
(see \cite{uzan,ekstrom} for a discussion of the problem
of calculating nuclear masses).
Nevertheless,
in this paper we will treat $\Qnic$ as a name for some
unknown combination of fundamental constants with the
dimension of energy.
We then write $\Qsn$ in a way that allows for TVOFC:
\begin{equation}
\Qsn(t_1) \sim \frac{\Qnic(t_1)}{56\alpha_G^{3/2}(t_1)}
\label{qsneq}
\end{equation}
Throughout this paper, 
$t_1$ will be the time of production of the photons that we
observe later at $t_0$.
We generally
show the argument $t_1$ only for quantities that
are \emph{not} expected to have any time dependence.
For things things that are expected to vary with time, like
the expansion rate $H(z)$, we generally use the redshift $z$ 
as the time parameter.

It is important to admit that equation \ref{qsneq} is a simplification
that does not do justice to the astrophysics that determines SNIa
luminosities.  The fact that the brightest and dimmest SNIa have
luminosities differing by a factor $\sim3$ demonstrates this
unfortunate fact.
This factor $\sim3$ is reduced to an effective $\Qsn$ dispersion
of $\sim12\%$ by using the observed durations and colors of
the supernovae (via the parameters $\alpha$ and $\beta$ in \cite{sndistances}).
In spite of these ``real world'' issues
we use (\ref{qsneq}) as the ``leading order'' effect
of the fundamental constants on SNIa luminosities.  
A lack of redshift evolution in the luminosity could then 
with some confidence 
be taken as evidence for the time-independence
of $\Qnic/\alpha_G^{3/2}$ multiplied by the factors that make
it dimensionless (see below).
Obviously, such evidence would only be as good as our understanding
of the astrophysics.

\subsection{Supernova in a box}

To see how the  constants impact
on  supernova measurements, we now need to model 
light propagation and the
detection process.
Before considering cosmological measurements, we
first consider a ``laboratory'' supernova where
the progenitor is placed
in an enormous cubic box that is able to resist the explosion.
The box, of volume $a^3$ has perfectly reflecting walls so
the photons emitted by the supernova are confined.
Later, the photons are counted and their wavelengths 
measured.
The total energy of the photons at the time of measurement ($t_0$) is 
\begin{equation}
E_{meas}=(hc)_{t_0}
\left\langle \frac{1}{\lsnm}\right\rangle\tilde{N}_\gamma
\end{equation}
where $\tilde{N}_\gamma$ and $\langle 1/\lsnm\rangle$ 
are the number of photons and their mean inverse wavelength.
Here, and throughout this paper, quantities with a tilde are quantities
measured at $t_0$.

If we consider the possibility of time-dependent fundamental constants,
we can't assume that the size of the box is time-independent.
If the changes are slow, this would cause the wavelengths of photons
to be adiabatically dilated by a factor $a(t_0)/a(t_1)$, corresponding
to a redshift, $z=a(t_0)/a(t_1)-1$.
If the box is constructed from solid material with a time-independent
number of atoms, then the size of the box is proportional to the
mean inter-atomic spacing.  
If the box is small enough to neglect gravitational tidal distortions,
it would be expected to have roughly
the same dependence on the fundamental constants as the Bohr radius, 
$a_B=\hbar/(\alpha m_ec)$.
We therefore write
\begin{equation}
\frac{a(t_1)}{a(t_0)}=
\frac{a_B(t_1)}{a_B(t_0)}=
\frac{(\hbar/\alpha m_ec)_{t_1}}{(\hbar/\alpha m_ec)_{t_0}}
\end{equation}
We can then deduce $\Qsn$ by modifying $E_{meas}$ to account for
the evolution of wavelengths and $hc$:
\begin{equation}
\Qsn(t_1)=(hc)_{t_1} \frac{a_B(t_0)}{a_B(t_1)}
\left\langle \frac{1}{\lsnm} \right\rangle\tilde{N}_\gamma 
=
2\pi(\alpha m_e c^2)_{t_1} a_B(t_0)    
\left\langle \frac{1}{\lsnm} \right\rangle
\tilde{N}_\gamma
\end{equation}
It is more instructive to write this equation as follows
\begin{equation}
\left(\frac{\Qsn}
{\alpha m_e c^2}\right)_{t_1} 
\;=\;
\left\langle \frac{2\pi}{\lsnm/a_B(t_0)} \right\rangle
\tilde{N}_\gamma
\label{dimensionstructure}
\end{equation}
The structure of this equation is very simple.
The l.h.s. is a dimensionless quantity that depends only
on fundamental constants at the moment of the explosion,
via equation \ref{qsneq}.
The r.h.s. is a product of a number of photons (counted at $t_0$)
and the dimensionless ratio of wavelengths ($\lsnm$)
and a length standard ($a_B$) all measured at $t_0$.
Measuring the r.h.s. for supernovas of different explosion
times then allows one to study the time evolution of the dimensionless
combination of fundamental constants given by
$\Qnic/\alpha_G^{3/2}\alpha m_ec^2$.
In fact, comparing directly the wavelengths of the two supernova
that exploded at times $t_1$ and $t_2$ and both detected at $t_0$
eliminates the need to measure them in units of $a_B$.
\begin{equation}
\frac
{\left(\Qnic/\alpha_G^{3/2}\alpha m_ec^2\right)_{t_1}}
{\left(\Qnic/\alpha_G^{3/2}\alpha m_ec^2\right)_{t_2}}
=\frac
{\tilde{N}_\gamma(1)\langle 1/\lsnm(1)\rangle}
{\tilde{N}_\gamma(2)\langle 1/\lsnm(2)\rangle}
\end{equation}
where the arguments $(1)$ and $(2)$ of the tilded quantities on the
r.h.s. refer to the measured values for the two supernovae.

Note that the reason we ended up being sensitive to dimensionless
constants at a given time comes from the fact that we 
assumed that $a(t)$ tracks a combination of fundamental constants,
that necessarily has the dimension of length 
and whose form
is determined by the ``experimental'' apparatus.  
In particular
we assumed $a(t)\propto a_B(t)$. 
If we had supposed $a(t)\propto 1/a_B(t)$, equation \ref{dimensionstructure}
would have become
\begin{equation}
\left(\frac{\Qsn}
{\hbar c a_B}\right)_{t_1} 
\;=\; a_B(t_0)^{-1}
\left\langle \frac{2\pi}{\lsnm} \right\rangle
\tilde{N}_\gamma
\hspace*{5mm}!!
\label{dimensionstructurewrong}
\end{equation}
The quantity on l.h.s. is a \emph{dimensionful} combination of fundamental
constants evaluated at $t_1$ (!).  
Taking the ratio of the r.h.s. for two SNIa of different $t_1$ but
the same $t_0$ would then allow us to follow the time evolution
of the dimensionful combination on the l.h.s.
But writing $a(t)\propto 1/a_B(t)$ requires us to define a new fundamental
constant $a_C$ (dimension of length) so that $a(t)=a_C^2/a_B(t)$.
Allowing for time variations of $a_C$ and defining $a_D(t)=a_C(t)^2/a_B(t)$
brings us back to the dimensionally correct form (\ref{dimensionstructure})
but with $a_B$ replaced by $a_D$.  We end up being only
sensitive to the dimensionless quantity, $\Qsn a_D/\hbar c$.

There is a second way  
to compare the two laboratory supernovas that does
not make any assumptions about the behavior of the box, and
is therefore more directly applicable to cosmological observations.
This method compares supernova photon wavelengths 
with the wavelengths of a known spectral line with the same redshift.
For example, we can use  the  \lya photons
($\lla=h/(3/4)\alpha^2m_ec$)
that are in principle identifiable in the spectrum of the
supernova or its host galaxy.
If they are measured 
at $t_0$
to have wavelength $\llam(z)$, 
we find
\begin{equation}
\Qsn(t_1)=\tilde{N}_\gamma \left\langle \frac{\llam(z)}{\lsnm} \right\rangle
\left[(3/4)\alpha^2m_ec^2\right]_{t_1}
\label{qsnrecon1}
\end{equation} 
It is better to write this in the form
\begin{equation}
\left(\frac{\Qsn}
{(3/4)\alpha^2 m_e c^2}\right)_{t_1} 
\;=\;
\tilde{N}_\gamma \left\langle \frac{\llam(z)}{\lsnm} \right\rangle
\label{qsnmc2}
\end{equation}
The combination of fundamental constants on the
l.h.s. 
is the total number of supernova photons that
would be produced if they were all \lya photons.
The factor
in $\langle\;\rangle$ on the r.h.s. corrects for the fact that
they are not.

Equation \ref{qsnmc2} shows that
by self-calibrating the detector with the SN \lya photons,
we find that rather than measuring $\Qsn$, we 
measure the dimensionless quantity $\Qsn/\alpha^2m_ec^2$.
By comparing such measurements from supernovae at different
ages, we can determine the time evolution of
$\Qnic/\alpha_G^{3/2}\alpha^2m_ec^2$.
One should note that the use of some other photon spectral line might
make one sensitive to a different combinations of fundamental
constants.  We use the \lya as an illustration because atomic
energy levels generally have $E\propto \alpha^2m_e c^2$ with
a constant of proportionality depending on the details of the atomic
shell structure.  As such, redshift determinations  using 
any atomic transition between states of differing principle quantum number 
would give limits
on $\Qnic/\alpha_G^{3/2}\alpha^2m_ec^2$.

The l.h.s.'s of equations \ref{dimensionstructure} and \ref{qsnmc2}
differ by a factor of $\alpha$ so
comparing results from the two methods
can give a direct determination of the time dependence
of $\alpha$.  One could also do this more directly 
by simply producing \lya photons at $t_1$, storing
them in a solid box of dimension $a$, and then comparing them
later at time $t_0$ with new \lya photons created at $t_0$.

\subsection{Cosmological supernovae}

\begin{figure}[tb] 
\epsfig{file=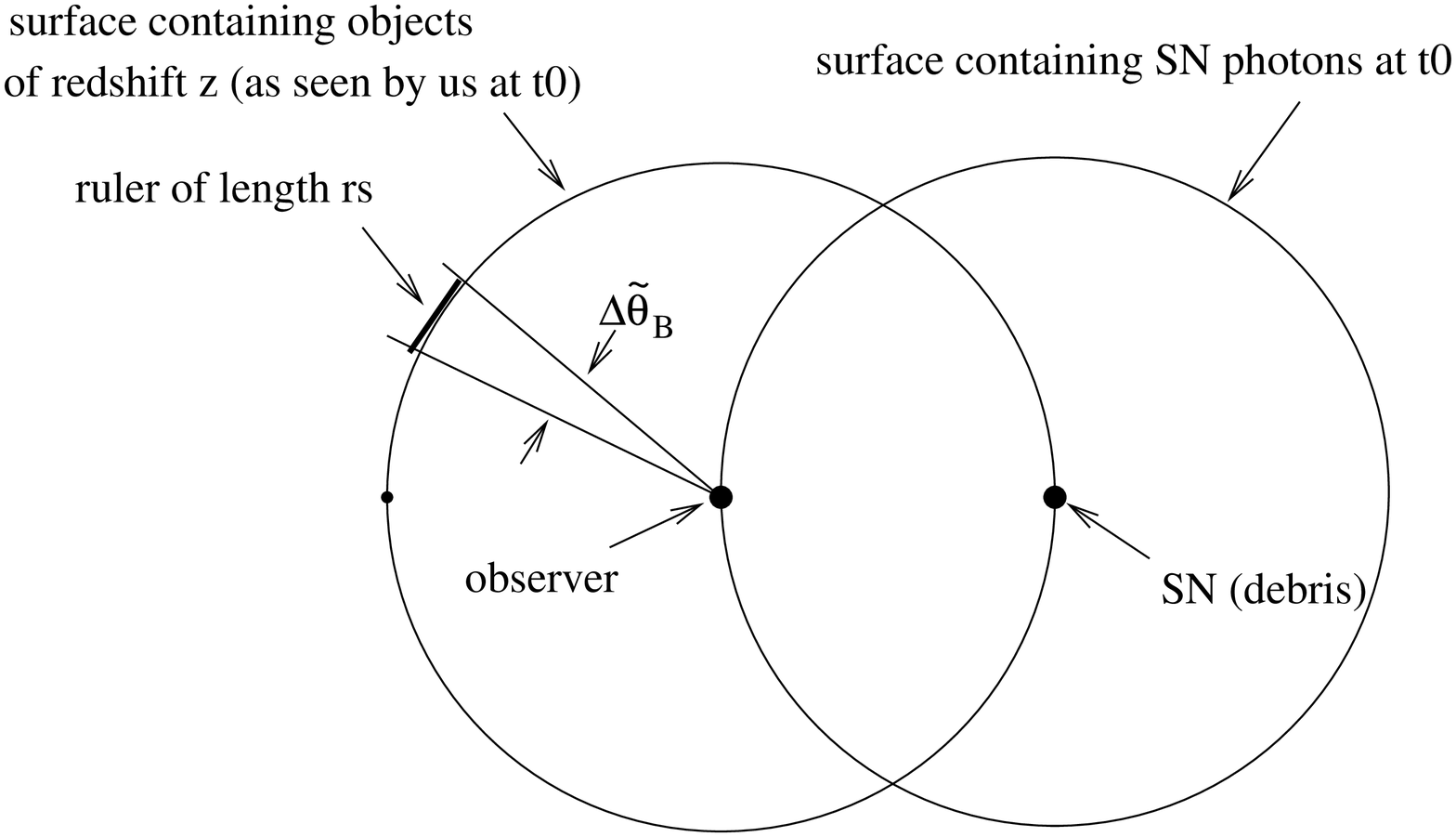, width = \textwidth} 
\caption{The universe at the time, $t_0$,  when
an observer detects the  photons from a supernova
at redshift $z$.
The figure shows the surface of two spherical shells, one centered 
on the supernova debris and the other on the observer.
The surface area of the first, $A(z)$, determines the fraction of the 
photons detected by the observer using a detector of area $\Adet$.  
The surface area of the second
is $4\pi r_s^2/\thetabao^2$ where $\thetabao$ is the
angle on the sky of the BAO standard ruler of length $r_s$.  
For a homogeneous universe, the two areas are equal.
Note that at the time of the explosion, $t_1$, the BAO ruler length was
smaller by the same factor as the sphere's radius, so the subtended angle
does not change.}
\label{t0surfacefig}
\end{figure}

We now attack real cosmological observations.  
Unlike the case of laboratory supernovae, we cannot
count all ($N_\gamma$) photons.
We count only the $\tilde{N}_\gamma$ photons that pass through a detector
of area $\Adet$: 
\begin{equation}
\tilde{N}_\gamma \,=\,
N_\gamma \,\frac{\Adet}{A(z)}
\label{measuredngamma}
\end{equation}
where $A(z)$ is the surface area of the sphere centered on the
point of the explosion and intersecting us at $t_0$, i.e. the size of
the detector that would be needed to detect all of the photons
(figure \ref{t0surfacefig}).
$A(z)$ is related to the usual ``luminosity distance'':
\begin{equation}
4\pi d_L(z)^2=(1+z)^2 A(z) 
\end{equation}
where the two factors of $(1+z)$ account for the redshift
of individual photons and cosmological time dilation.

As for laboratory supernovae, the total number of photons
is $\Qsn/(3/4)\alpha^2m_ec^2$,  evaluated at $t_1$ and corrected
for the fact that the mean photon energy is not equal to that
of a \lya photon.  Equation \ref{measuredngamma} then becomes
\begin{equation}
\tilde{N}_\gamma \,=\,
\left(\frac{\Qsn}
{(3/4)\alpha^2 m_e c^2}\right)_{t_1} 
\left\langle \frac{\llam(z)}{\lsnm} \right\rangle^{-1}
\,\frac{\Adet}{A(z)}
\end{equation}
or equivalently
\begin{equation}
A(z)\,=\,\Adet
\left( \frac{\Qsn}{(3/4)\alpha^2m_ec^2}\right)_{t_1}
\left[
\left\langle
\frac{ \llam(z) }{\lsnm(z)}
\right\rangle 
\right]^{-1}
\,\tilde{N}_\gamma^{-1} 
\label{asneq}
\end{equation}
This is equivalent to the standard relation between $\Qsn$, the
photon flux, and the luminosity distance.

Before going on to BAO, we write equation \ref{asneq}
in terms of the ``published'' areas, $A_{\rm SN0}(z)$, that are derived
from SNIa fluxes assuming no TVOFC that would lead to a spurious
redshift dependence of $\Qsn$:
\begin{equation}
A(z)= A_{\rm SN0}(z)\frac
{\left[\Qnic/(\alpha_G^{3/2}\alpha^2m_ec^2)\right]_{t_1}}
{\left[\Qnic/(\alpha_G^{3/2}\alpha^2m_ec^2)\right]_{t_0}}
\label{asneq2}
\end{equation}
Again, we assume that 
the primary dependence of $\Qsn$ on  the fundamental constants
is contained in the simple formula (\ref{qsneq}).

\subsection{BAO}

Baryon Acoustic Oscillations took place 
before electron-proton recombination
(redshifts $z>z_{rec}\sim1070$)
when the universe supported acoustic waves
in the electron-proton-photon plasma.
This plasma  was a nearly
perfect fluid because of Compton scattering of the photons on the
free electrons.
Sound waves propagating out of initial perturbations had the
effect of separating the baryon-electron-photon component from
the cold-dark-matter component which, being collisionless,  did
not generate waves.
This separation is seen now as
an enhancement in the  two-point matter
correlation function at a (co-moving) distance equal to the
the ``sound horizon'', i.e., the distance that a sound wave
could travel between the big-bang and 
recombination
\begin{equation}
r_s \,= \int_{z_{rec}}^\infty \frac{c_s(z) dz}{H(z)} \,\sim 150~{\rm Mpc}
\end{equation}
Here, $H(z)$ is the expansion rate and $c_s(z)$ is the speed of sound.
The excess correlation at this distance is ``co-moving'' in the
sense that the physical distance expands with the Universe.
This means that 
the  excess correlation is for points separated by  $r_s/(1+z)$
where $z$ is the  redshift.
This expansion of the BAO ruler by the factor $1+z$ is expected
to be precise at the 1\% level.
We will not need to know the actual value of $r_s$ since it will
cancel out in applications for TVOFC.

Seen on the sky, the distance of enhanced correlations corresponds
to angular separations and redshift separations of
\begin{equation}
\thetabao \,=\, \frac{r_s}{D(z)}
\hspace*{10mm}
\dzbao\, =\, r_s H(z)
\end{equation}
where $D(z)$ is the co-moving angular distance to the redshift $z$  
(equal to the usual angular distance, $D_A(z)$, multiplied by $(1+z)$)
and $H(z)$ is the expansion rate at redshift $z$.
The first relation is the simple geometric relation illustrated
in figure \ref{t0surfacefig}.
The second is a bit more subtle and we will derive in later
(equation \ref{hbao1}).

The angular BAO effect gives a relation between the 
angular separation, $\Delta\tilde{\theta}$ on the sky of two points 
at a common redshift $z$
and the physical distance, $d$, between the two points (at the moment of observation):
\begin{equation}
d=r_s \frac{\Delta\tilde{\theta}}{\thetabao(z)}
\end{equation}
This gives the area of the sphere surrounding us and
intersecting the supernova debris:
\begin{equation}
A(z) = \frac{4\pi r_s^2}{\thetabao(z)^2}
\label{abao}
\end{equation}
As figure \ref{t0surfacefig} shows,
this is not the same sphere as the one that determines
the supernova flux but in a homogeneous universe their
surface areas must be the same, because they have the same radius.
The equivalence of the two areas in a homogeneous universe
is a special case of the more general relation between luminosity
and angular distances, $d_L(z)=(1+z)^2d_A(z)$ \cite{lumangdistref}.

Equation \ref{abao}
has the same structure of equation \ref{asneq} except that
there is no (dimensionless) combination of fundamental constants.
We note that the dimensioning role of $\Adet$ for supernovae as been taken
by $r_s^2$ for BAO.

\begin{figure}[tb] 
\epsfig{file=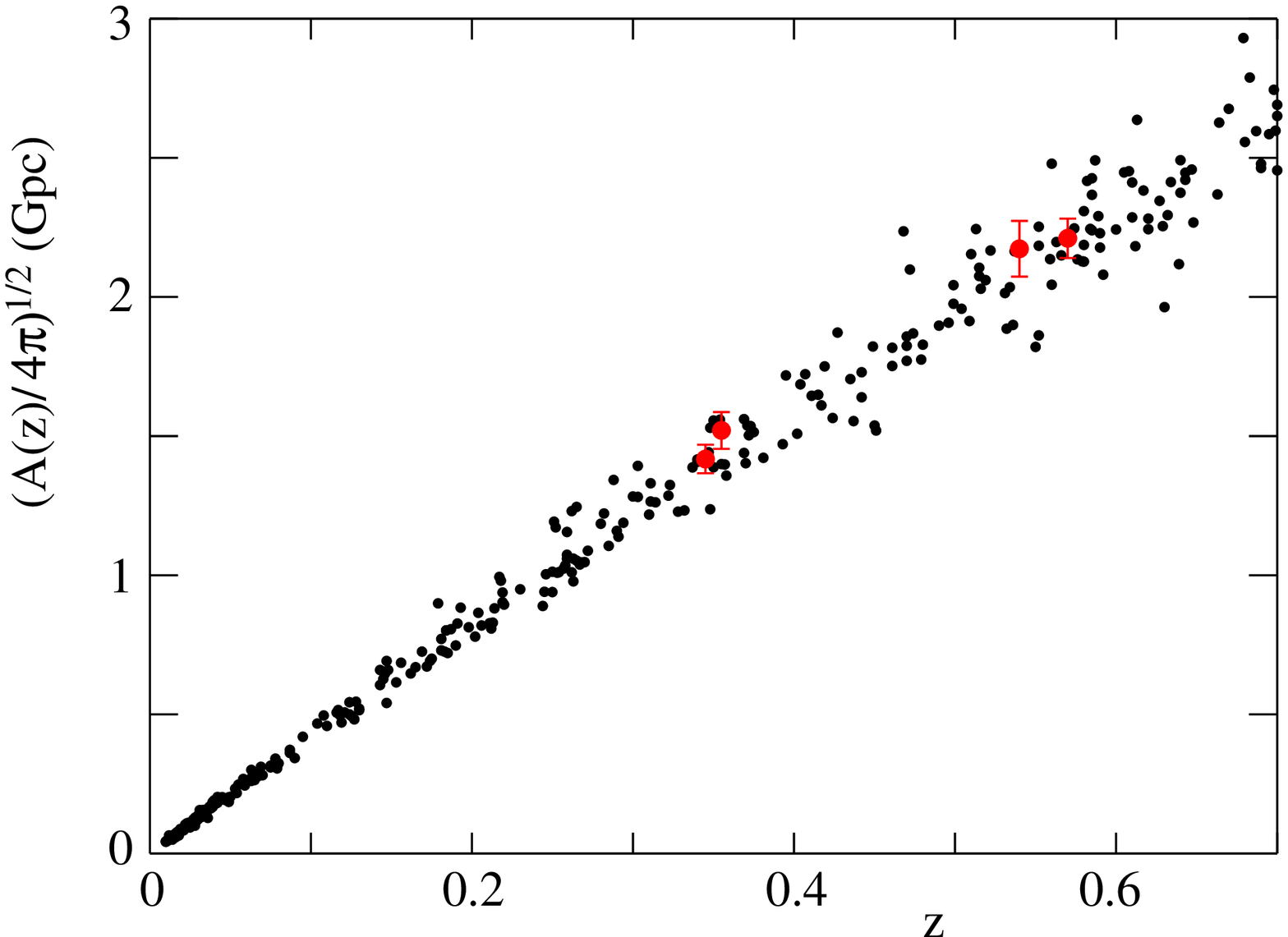, width = \textwidth} 
\caption{Measurements of $A(z)\,vs.\,z$.
The black dots are individual SNIa from the compilation of \cite{sndistances}.
The red dots are 
BAO measurements at $z=0.35$ \cite{Chuangref,Xuref},
$z=0.54$ \cite{Seoref}.
and $z=0.57$ \cite{Anderson2013}.
The BAO measurements assume $r_s=152{\rm Mpc}$. The
supernovae distances are normalized so that $H_0=70~{\rm km\,s^{-1}Mpc^{-1}}$
and use the stretch and color parameters $\alpha=1.4$ and $\beta=3.15$.
}
\label{hubblefig}
\end{figure}

For the time being, there are only
four published BAO measurements of $A(z)$ 
\cite{Chuangref,Seoref,Xuref,Anderson2013}
based on the correlation function of galaxies.
They are shown in
figure \ref{hubblefig} along with individual SNIa measurements
of $A(z)$ taken from \cite{sndistances}. 
There is also a new BAO measurement of $A(z=2.3)$ using the flux
correlation function in the Ly$\alpha$ forest 
\cite{anze}
but this
is at too high a redshift to be compared directly with
SNIa measurements.

\subsection{Comparing two distance measurements}

To set limits on time variations of fundamental constants
we need only equate the area  measurement using SNIa
(equation \ref{asneq2}) with  that from BAO (equation \ref{abao}):
\begin{equation}
\frac{4\pi r_s^2}{\thetabao(z)^2}
\,=\,  A_{\rm SN0}(z)\frac
{\left[\Qnic/(\alpha_G^{3/2}\alpha^2m_ec^2)\right]_{t_1}}
{\left[\Qnic/(\alpha_G^{3/2}\alpha^2m_ec^2)\right]_{t_0}}
\label{twoareas}
\end{equation}
If the BAO $A(z)$ on the l.h.s. is equal to $A_{\rm SN0}(z)$
on the r.h.s. we could conclude that $\Qnic/(\alpha_G^{3/2}\alpha^2m_ec^2)$
has not varied significantly between $t_1$ and $t_0$.
Such a conclusion based on the
BAO and SNIa data in figure \ref{hubblefig}
would, however, depend on the ``normalization'' of the BAO and SNIa
distances.
For the BAO, this amounts to using the value of $r_s$ derived from
CMB data.
For the SNIa data, it requires knowledge of the absolute value
of the mean $\Qsn$ or, equivalently, knowledge of the Hubble
constant.
(The Hubble constant, $H_0$, can be used to normalize the
distances since for $z\ll 1$ we must have $A(z)=4\pi(zc/H_0)^2$.)
It is safer to eliminate the dependence on
BAO and SNIa normalizations by taking
ratios of $A(z)$ at different redshifts.
For the BAO measurements we find
\begin{equation}
\frac{A(z=0.57)}{A(z=0.35)} = 2.26\pm0.26 \; \hspace*{5mm}{\rm BAO} .
\end{equation}
Averaging the SNIa data over the redshift ranges $0.3<z<0.4$
and $0.51<z<0.61$ 
gives 
\begin{equation}
\frac{A_{\rm SN0}(z=0.57)}{A_{\rm SN0}(z=0.35)} 
\,= 2.40\pm0.10 \; \hspace*{5mm}{\rm SNIa} .
\end{equation}
Injecting these two results into (\ref{twoareas}) we find
\begin{equation}
\frac
{\left[\Qnic/(\alpha_G^{3/2}\alpha^2m_ec^2)\right]_{t(z=0.57)}}
{\left[\Qnic/(\alpha_G^{3/2}\alpha^2m_ec^2)\right]_{t(z=0.35)}}
\,=\,
0.94 \pm 0.12
\end{equation}
i.e. less than $\sim10\%$ variation between the two redshifts.
The time interval between $z=0.57$ and $z=0.35$ is 
$\sim1.5\times10^9~{\rm yr}$ for the standard cosmological
parameters.
The upper limit on the logarithmic time
derivative of $\Qnic/(\alpha_G^{3/2}\alpha^2m_ec^2)$ 
is therefore $\sim10^{-10}{\rm yr^{-1}}$.  This  is considerably weaker than
other published limits on TVOFC.
For example, limits on $\dot{G}/G$ from lunar ranging
data are in the range
$\sim10^{-12}{\rm yr^{-1}}$ \cite{lunarranging}.
Our effort has, however,  given us the satisfaction of
having a limit on a dimensionless combination of constants
that includes, to first order, all the essential physics.
Our limit also refers to
time variations at a different mean time and in different
regions of space than the lunar-ranging limit.

\subsection{Fundamental rulers}

It is instructive to replace the BAO standard ruler (whose size
expands with the universe) with a hypothetical solid ruler
whose length is some known number, $\tilde{N}_x$, times the Bohr radius.
In this case, the area of equation \ref{abao} is replaced
with
\begin{equation}
A(z) = \frac{4\pi (\tilde{N}_xa_B(t_1))^2}{\tilde{\theta}_x(z)^2} (1+z)^2
\label{afixed1}
\end{equation}
where $\tilde{\theta}_x$ is the observed angle subtended by
the ruler on the sky. Two things are important:  first
we use $a_B(t_1)$ instead of $a_B(t_0)$ 
because it is the ruler's size when it
emits photons that determines it's angle on the sky.
Second, compared to (\ref{abao}) we have two factors of
$(1+z)$ that reflect the increase in the area of the sphere's surface
between $t_1$ and $t_0$.

The redshift is given by the ratio between emitted and received
\lya photons
\begin{equation}
1+z=\frac{\llam(z) }{ [h/ (3/4)\alpha^2m_ec]_{t_1} }
\label{redshifteq}
\end{equation}
This gives
\begin{equation}
A(z)\,=\,\llam^2(z)\, 
\frac{\tilde{N}_x^2 }{\tilde{\theta}_x(z)^2}\, 
\alpha(t_1)^2 \;\frac{9}{16\pi}\,
\label{afixed2}
\end{equation}
This now has the same dimensional structure as equations \ref{abao} and
\ref{asneq}: products of local measurables (with a total dimension of
length-squared)
with a dimensionless
combination of fundamental constants at $t_1$.
Comparing areas measured with (\ref{afixed2}) with those measured
with (\ref{abao}) would allow one to study the time dependence of $\alpha$.  
Comparing them with the SN-derived area (\ref{asneq}) would allow one
to study the  time dependence of
the dimensionless combination $\Qnic/\alpha_G^{3/2}\alpha^4m_ec^2$.

\subsection{Why only dimensionless combinations?}

We now have three measurements of $A(z)$ using standard candles (equation
\ref{asneq}), co-moving standard rulers (\ref{abao}) and 
fixed standard rulers (\ref{afixed2}).  For all three, the
area depends on a \emph{dimensionless} combination of fundamental constants
at $t_1$.  Taking the ratio of any two measurements of $A(z)$ then
allows us to follow the time evolution of a dimensionless combination.
Why is this?  Why, for instance, didn't we find a method
that gives the dimensionally correct area
\begin{equation}
A(z) = \frac{4\pi \tilde{N}_x^2}{\tilde{\theta}_x(z)^2} 
a_B^2(t_1)
\hspace*{5mm}!!
\label{acrazy}
\end{equation}
i.e. equation \ref{afixed1} without the two factors of $(1+z)$.
Taking the ratio of this area with any of the other three would
allow us to track the time evolution of a dimensionful quantity proportional
to $a_B^2$.

While (\ref{acrazy}) is dimensionally correct, it makes no physical
sense because it does not take into account the expansion
of the universe.
The three correct methods of measuring $A(z)$ all do this in their own way,
leading to sensitivity  to a dimensionless combination.
For standard candles, the dimensional factor for $A(z)$ is
given by $\Adet$.  The fundamental constants only enter because
they determine the (dimensionless) number of photons produced
through the dimensionless combination $\Qsn/\alpha^2m_ec^2$.
This number is then corrected by the factor 
$\llam/\langle\tilde{\lambda}\rangle$ which assumes that all
photon wavelengths evolve with the same scale factor

For the two methods using standard rulers, the dimensional factor
for $A(z)$ is the square of the ruler length.  More precisely, 
it is the ruler length at $t_1$ (which determines the angular size
of the ruler) scaled up by a factor $(1+z)$ to take into account
expansion between $t_1$ and $t_0$.
For the BAO co-moving ruler and the solid ruler the length factors
are
\begin{equation}
\frac{r_s}{1+z}\times (1+z)
\hspace*{10mm}
a_B(t_1)(1+z)
\end{equation}
For co-moving rulers, the factors of $(1+z)$ cancel and we end
up with $A(z)$ being independent of the fundamental constants, i.e.
dependent
on a particularly simple dimensionless combination.
For fixed rulers, the assumption that the expansion factor $1+z$
is given by a spectroscopic redshift given by equation  \ref{redshifteq}
allows us to shift the dimensionality 
from $a_B(t_1)$ to the measured $\llam$, leaving
us sensitive to a dimensionless combination of constants
evaluated at $t_1$.

The fact that the ratio of areas given by any combination of
two methods thus depends on the existence of a unique
scale factor that governs the expansion of all photon 
wavelengths as well as the distances between co-moving points.

\section{The expansion rate}

We now discuss measurements of TVOFC using the expansion
rate at a redshift $z$.
Again, we expect to be sensitive only to dimensionless
combinations of constants, but, as usual, it is not
obvious how this comes about.

\begin{figure}[tb] 
\epsfig{file=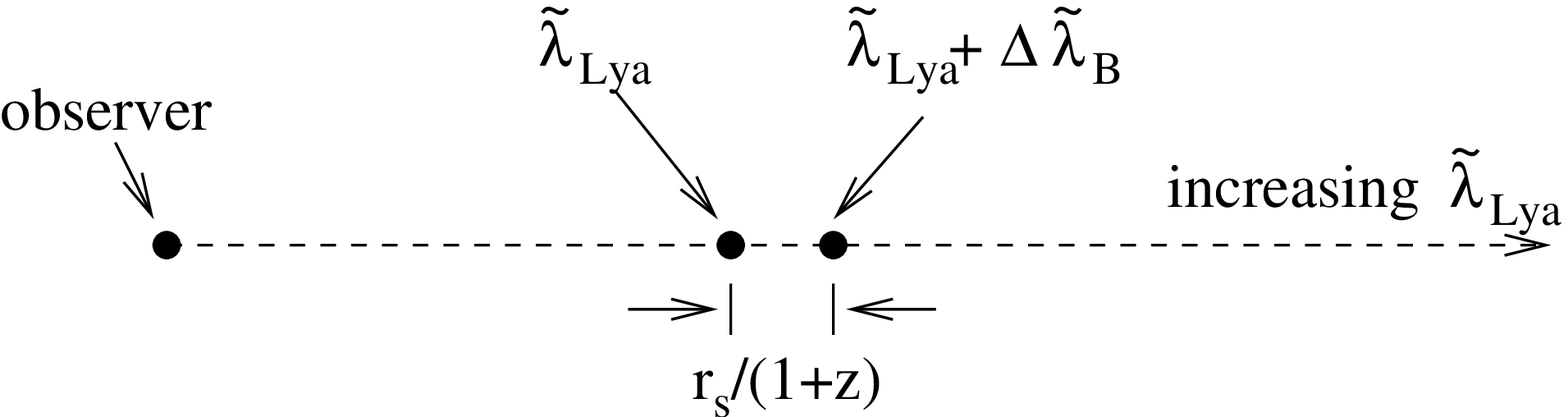, width = \textwidth} 
\caption{Two points in the same direction on the sky that
are seen by the observer to have \lya emission at $\llam$
and $\llam+\dlbao$.  
An observer at the  point $\llam$ will see photons emerging from the
point $\llam+\dlbao$ Doppler shifted by $\dlbao/\llam$.
At the time, $t_1$, of emission of the photons seen by the observer, 
the two points are separated by a distance $r_s/(1+z)$.
Hubble's law then implies an expansion rate given by equation (\ref{hbao1}). }
\label{rsshellfig}
\end{figure}

Other than $z=0$ measurements of $H_0$, the most direct
measurements \cite{Chuangref,Xuref,Anderson2013,anze,busca} use the 
BAO peak in the radial (redshift) direction.
For galaxies, we will use as a radial coordinate
the wavelength of its \lya photons, $\llam$, related to the
redshift via equation \ref{redshifteq}.
Neglecting proper motions,  $\llam$
increases monotonically with distance from the observer,
as illustrated in figure \ref{rsshellfig}.
The mass correlation function plotted in this variable
has a peak at $\llam$ separations equal to $\dlbao$,
corresponding to the co-moving
sonic horizon, $r_s/(1+z)$.
The wavelength separation of the BAO peak is
related to the expansion rate, $H(z)$ that would be measured
by an observer at that redshift:
\begin{equation}
H(z) = \frac{c(t_1)\dlbao(z)/\llam(z)}{r_s/(1+z)}
\label{hbao1}
\end{equation}
To understand this relation, we note that
the numerator of the r.h.s. is
the Doppler shift that an observer at the point $\llam$ would see
for the photons coming from the point $\llam+\dlbao$.
The denominator is the distance between the 
two points.
By Hubble's law, the ratio gives the expansion rate, i.e. $H(z)$
on the l.h.s.

Using (\ref{redshifteq}) we get
\begin{equation}
H(z) 
= \frac{\dlbao(z)}{r_s}
\left( \frac{ (3/4)\alpha^2m_ec^2 }
            {h}            \right)_{t_1} 
\label{hbaoeq}
\end{equation}
We see that $H(z)$ as  determined by a dimensionless
combination of local measurables $(\dlbao/r_s)$ and a 
combination of fundamental constants at $t_1$ with the dimension of
1/time.
It is interesting that the dimensional factor is established at $t_1$
while for measurements of $A(z)$ the dimensionality is
established at $t_0$ (i.e. factors of $\Adet$, $r_s^2$, $\llam^2$).
This might not be surprising because $H(z)$ refers to the expansion
rate at $t_1$ while $A(z)$ is the area of the surface at $t_0$.

Equation (\ref{hbao1}) suggests a more general technique
for measuring $H(z)$ by using a clock to determine the
time separation $\Delta t$ between the instants that 
we see two points in the same direction at redshifts corresponding to
$\llam$ and $\llam + \Delta\llam$.
\begin{equation}
H(z)\,=\, \frac{1}{\Delta t}
\frac{\Delta\llam}{\llam}
\label{hgeneral}
\end{equation}
This is found by using $c\Delta t=r_s/(1+z)$ in (\ref{hbao1}).

To use equation \ref{hgeneral} to find $H(z)$, we need clocks
that tell us how cosmic time changes with redshift.
The evolution of stellar populations has been used to do this
\cite{moresco} and the measurements of $H(z)$
agree with BAO measurements at the 10\% level over
the range $0<z<2$.
However, to extract clear limits on TVOFC, the $H(z)$
from stellar chronometers 
needs to be put in a form analogous to that
of equation \ref{hbaoeq} where the role of the fundamental 
constants is clear.  
The results of \cite{moresco}
are not easily put in this form.
Here, we will only illustrate the basic idea by
considering a simpler clock based on the lifetime
of main-sequence stars.
Such stars have a lifetimes that increase like roughly
the third power of the mass of the star.
If all stars were created at a unique time, then the
time corresponding to any redshift is then simply the
lifetime of the heaviest remaining star.

To see what fundamental constants determine a stars lifetime,
we follow the simple model of \cite{weisskopf}.
For a star starting it's life with $N_n$ nucleons, it's
inverse lifetime is
\begin{equation}
T_{ms}^{-1}=\frac{L}{N_n f \QH}
\end{equation}
where $L$ is it's luminosity (averaged over it's lifetime),
$4\QH$ is the energy released in the
transformation of four hydrogen atoms  to helium,
and $f$ is the fraction of the star's nucleons that 
is available for burning (only protons near the center  burn).
The luminosity is given by
\begin{equation}
L = 
\frac{E_\gamma}{\tau_\gamma}
\,\sim\,
\frac{G^4N_n^3m_p^8c}{(\hbar c)^3\sigma}
\end{equation}
where $E_\gamma$ is the total energy of the black-body radiation
inside the star and $\tau_\gamma$ is the mean time for a black-body
photon to escape from the star.
The second form \cite{weisskopf} uses hydrostatic equilibrium
to determine the mass-temperature relation and photon random walks
to determine $\tau_\gamma$ in terms of the effective photon cross section
$\sigma$.
(Given the approximations in the model, 
we now drop numerical factors and use $\sim$.)
The inverse of the main-sequence lifetime is then
\begin{equation}
T_{ms}^{-1} \sim
\frac{G^4N_n^2m_p^8c}{(\hbar c)^3\sigma f\QH}
\end{equation}
Since $N_n$ is not directly measurable for a star, we prefer to write
the this as
\begin{equation}
T_{ms}^{-1} \sim
\left(\frac{GN_nm_p}{c^2r_s/(1+z)}\right)^2 [r_s/(1+z)]^2
\frac{\alpha_G^2 (m_pc^2)^2 }{f\QH \sigma\hbar}
\end{equation}
The term in parentheses is the gravitational potential (divided by $c^2$)
at a distance of $r_s/(1+z)$ from the star.  It can be measured by observing
the velocity dispersion of objects orbiting around the star and
separated from it by an angle $\tilde{\theta}$:
\begin{equation}
\tilde{\Phi}\equiv  \frac{GN_nm_p}{c^2r_s/(1+z)}
\,=\,
\frac{v^2}{c^2}\frac{\tilde{\theta}}{\thetabao}
\,=\,
\left(\frac{\Delta\tilde{\lambda}}{\tilde{\lambda}}\right)^2
\frac{\tilde{\theta}}{\thetabao}
\end{equation}
where $\tilde{\lambda}$ refers to the wavelength of any atomic
line that can be used to measure the orbital velocity dispersion
$\Delta\tilde{\lambda}/\tilde{\lambda}$.

Using the expression for the redshift we have
\begin{equation}
T_{ms}^{-1} \sim
\tilde{\Phi}^2 \, \left(\frac{r_s}{\llam(z)}\right)^2
\frac{\alpha_G^2 (m_p/m_e)^2 }{(3/4)^2\alpha^4 \sigma f\QH} \hbar c^2
\end{equation}
For heavy stars $M>5M_\odot$, photon diffusion is dominated by
scattering on free electrons. For such stars we can
take $\sigma=\sigma_t\sim\alpha^2(\hbar/m_ec)^2$ giving
\begin{equation}
T_{ms}^{-1} 
\sim \tilde{\Phi}^2 \, \left(\frac{r_s}{\llam(z)}\right)^2
\frac{\alpha_G^2 (m_p/m_e)^2 (m_ec^2)^2}{(3/4)^2\alpha^6  f\QH}  \hbar^{-1}
\end{equation}

In principle we can then measure time
intervals in units of $T_{ms}$ by following
the redshift evolution of stellar populations.
We then  set $\Delta t=\tilde{N}_tT_{ms}$ 
in equation \ref{hgeneral} giving
\begin{equation}
H(z)\sim\frac{\Delta\llam}{\langle\llam\rangle}\,
\tilde{\Phi}^2 \, \left(\frac{r_s}{\llam(z)}\right)^2
\tilde{N}_t^{-1}\,
\left(
\frac{\alpha_G^2 (m_p/m_e)^2 (m_ec^2)^2}{(3/4)^2\alpha^6  f\QH}  \hbar^{-1}
\right)_{t_1}
\label{hmseq}
\end{equation}
This now has the same form as equation \ref{hbaoeq}:
the product of a dimensionless combination of local measurables and a 
combination of fundamental
constants evaluated at $t_1$ with the dimension 1/time.
Setting the two measurements of $H(z)$ equal to each other,
we can investigate the time dependence of the ratio of the
two dimensioned combinations, i.e. the dimensionless combination
\begin{equation}
\frac{\alpha_G^2 (m_p/m_e)^2 m_ec^2}{\alpha^8  f\QH}  
\end{equation}

The fact that we ended up investigating the time dependence
of a \emph{dimensionless} combination of constants
is due to the fact that both methods for determining $H(z)$ 
are sensitive to a combination of constants with the 
dimension of inverse time.
For the second method (\ref{hmseq}) this comes about
because we use equation \ref{hgeneral} with the clock
period depending only on a combination of constants
with the dimension of time.
For the first method (\ref{hbaoeq}), we used a BAO ruler
whose length is assumed to follow the same expansion law
as that for photon wavelengths,  transforming
the $c(t_1)$ dependence in equation \ref{hbao1} into the
properly dimensioned equation \ref{hbaoeq}.
A third hypothetical method that used a ruler fixed by fundamental
constants would have given an inverse time dependence
directly proportional to $c(t_1)$ divided by the ruler length,
e.g. $\propto c(t_1)/a_B(t_1)$.

\section{Conclusion}

Cosmology is largely based upon measurements of photon fluxes,
angles on the sky, and redshifts.
These measurements are then combined to yield the
two fundamental cosmological observables:  distances and expansion rates. 
Individual measurements depend on cosmological parameters and
fundamental constants but information on the latter can be
extracted by comparing pairs of redundant measurements.
We have shown that pairs of  distance measurements using SNIa and BAO and 
expansion rate measurements using BAO and stellar chronometers
are sensitive only  to dimensionless 
combinations of fundamental constants.
This creates a challenge for advocates 
``Varying Speed of Light'' models \cite{Moffat}.
These theories have been widely criticized for
a variety of reasons \cite{ellisuzan}.
Here, we  suggest only that
theories where  $c(t)$ appears in equations should
use a different name until they
describe a  measurement that can directly measure 
how the speed of light varies (with time).
This criticism does not apply to ``$c(\lambda)$'' theories
where the speeds of photons of different wavelengths can
directly give the wavelength dependence of the speed of light.

We emphasize that
the reason that distance and expansion-rate measurements
are sensitive only to dimensionless constants comes from 
the fact that useful cosmological rulers
have lengths that are either fixed by fundamental
constants or expand in the same way as photon wavelengths.
These are the most trivial ways for lengths to evolve.
There are, of course rulers that have more complicated
time evolutions.
For example, stellar radii
slowly change with time  as a star evolves.
But this makes such rulers dependent not only on the
fundamental constants, but also on the age of the star.
As such, they not very useful for looking for TVOFC.
On a more fundamental level, one could imagine laws
in which the distances between co-moving observers
do not follow the same expansion law as photon
wavelengths.  
This would violate a rather  straightforward prediction of 
general relativity that is essentially equivalent to an integration of the
non-relativistic Doppler shift of photon wavelengths.
As such, changing the standard law for photon redshifts 
would probably involve a more fundamental
change in physical law than just allowing the constants
to vary in time.

\section*{Acknowledgments}
It is a pleasure to thank Jean-Philippe Uzan for discussions on
fundamental constants and Nicolas Busca, John LoSecco and Graziano
Rossi for comments on the text.

\end{document}